# An Empirical Study of User Support Tools in Open Source Software

Arif Raza, Luiz Fernando Capretz, *Senior Member, IEEE*, and Shuib B. Basri

*Abstract*—End users' positive response is essential for the success of any software. This is true for both commercial and Open Source Software (OSS). OSS is popular not only because of its availability, which is usually free but due to the user support it provides, generally through public platforms. The study model of this research establishes a relationship between OSS user support and available support tools. To conduct this research, we used a dataset of 100 OSS projects in different categories and examined five user support tools provided by different OSS projects. The results show that project trackers, user mailing lists, and updated versions have a significant role in gaining user support. However, we were unable to find a significant association between user support and documentation, as well as between user support and the troubleshooting guidelines provided by OSS projects.

## I. INTRODUCTION

One of the main priorities of the software industry is to develop high quality and efficient software [1]. This is true for both commercial and Open Source Software (OSS). OSS promotes volunteer collaboration in a distributed and decentralized environment resulting in a relatively lower production cost and improved software quality [2]. Easy access to the Internet has enhanced user involvement and frequent downloads of OSS.

Raza et al. [3] observe that as the popularity of OSS grows, OSS community, made up of individuals and organizations, has also extended its boundaries. OSS is not limited to technically adept individuals; novice users use many open source applications as well. This brings a new set of challenges for OSS developers in the form of enhanced user requirements and support. The authors also show concerns for the effectiveness of public forums in the OSS environment.

In comparison to commercial proprietary software, OSS faces a more challenging environment as far as user support is concerned [4, 5]. Many OSS users who are not necessarily technically adept persons have varied backgrounds in terms of education, culture, needs, and requirements. Bodker et al. [6] state that OSS products are not so user-friendly. The authors observe that this shortcoming is mainly due to the lack of focus on user expectations by OSS developers and their teams. According to Nichols and Twidale [7], the prime reason for this weakness in the OSS environment is the distinction between the users and the developers.

A. Raza is with School of Engineering, Technology and Trades, Confederation College, Thunder Bay, Ontario, Canada – P7C 4W1 (e-mail: araza@confederatiocollege.ca).
L.F. Capretz is with the Department of Electrical & Computer Engineering, Western University, London, Ontario, Canada – N6A 5B9 (e-mail:lcapretz@uwo.ca), URL: www.eng.uwo.ca/electrical/faculty/capretz_l
S.B. Basri is with the Computer and Information Sciences Department, Universiti Teknologi PETRONAS, Bandar Seri Iskandar, Malaysia – 32610 (e-mail: shuib_basri@utp.edu.my).

In this study, we have examined five user support tools provided by different OSS projects. In order to carry out this analysis, we studied the relationship between the number of weekly downloads and the availability of different user support tools namely project trackers, troubleshooting guidelines, user mailing lists, documentation, and updated versions. A dataset of 100 OSS projects covering a varied range of categories has been used to empirically examine the study model for this research.

The rest of this paper is organized as follows: the next section presents the literature review that inspired the authors to carry out this research study. The model and research questions are explained in Section III. The data collection process and the research methodology are explained in Sections IV and V. Finally, discussions of results and conclusion are described in Sections VI and Section VII respectively.

## II. OPEN SOURCE SOFTWARE AND USER SUPPORT

While focusing on users' requirements in the OSS domain, Raza and Capretz [8, 9] wonder whether OSS developers listen to their users. Whereas Balter [10] states that *"Software is written by humans for humans."* Referring to open source, he maintains that there cannot be a "litmus test for 'good' open source," however, there are certain factors which may be considered as "indicators of a healthy project." On the other hand, Crowston et al. [11] maintain that although OSS is an innovative approach, it is now considered an integral part of the modern software industry. It is a known fact that, in general, any developer contributions to OSS is voluntary. The authors mention on the basis of their analysis: "The literature to date has been relatively limited in scope and many aspects of FLOSS development have received a little examination." (FLOSS is an acronym for free/libre open source software). Hence, there is a need for researchers to understand and effectively manage software development in the OSS world.

Von Krogh et al. [12] agree that OSS development differs from conventional software development. They, however, highlight three dimensions, "incentives, control and coordination mechanisms," in particular. They consider OSS as a relatively new concept that has its own motivation to contribute to the development of information systems. It is known that OSS contributors include both paid developers and unpaid volunteers; these authors emphasize that even in OSS, "standards of excellence emerging in global communities of software developers can gain broad endorsement and impact quality standards expected by users and customers."

Viorres et al. [13] consider educational reasons, re-usability, and developing a reputation as some of the valid reasons to adopt OSS. However, they are concerned about certain issues in OSS such as usability, installation and maintenance, backward compatibility, and documentation. Whereas Hedberg et al. [14] propose that in order to achieve a higher quality and address user-related issues, proven methods need to be adapted in OSS as well. Similarly, regarding user involvement, Iivari and Iivari [15] realize that in a real world it is not possible to access every potential software user to understand their demands, thus they need to be trained to have an adaptable and compatible system. However, according to Garzarelli et al. [16], OSS claims its share favorably in a market as compared to proprietary commercial software mainly due to its licensing policies. Thanks to its volunteer contributors, OSS flourishes and allows sharing of its code and standards.

Although many consider user satisfaction to be a prime factor in any software success, whether proprietary and OSS [17, 18, and 19], there is a lack of empirical studies addressing the vital issues being faced by OSS communities [20]. A large population of OSS users are software developers themselves; still, different OSS user communities have their own motivations and requirements. Tiemann [21] observes that the traditional approach to software has changed dramatically with the emergence of OSS. Developers and users are considered to be "a creative continuum, not a distinct caste" in the OSS world. The author also maintains that "the sourceforge.net web site has equaled or exceeded Microsoft's productive potential using a social, not an industrial model." Dahlander and Mckelvey [22] observe that different people and even organizations may have their own motives in contributing to OSS. The authors believe that OSS should not rely solely on the contribution of its development force, i.e., core developers. They suggest that in order to achieve success, OSS users also need to actively play their role in OSS development.

Anyhow, the boundary of the open source community is expanding day by day. OSS users are not limited to technical gurus anymore. As observed by Smaja [23], people such as political activists, business professionals, and commentators are using OSS for all sort of social activities. This confirms their statement that "there is more to open source than a vague commitment to transparency, participation, and collaboration."

### III. STUDY MODEL AND PROPOSITIONS

The progression of OSS projects and growth in the number of its users has expanded enormously in the past decade. In the OSS environment, there are different platforms and tools available for diverse contributors to share their thoughts. These tools play an active role in managing these projects. The main drive of this study is to explore the answer to the following research question (RQ):

RQ: How much OSS projects support the users?

The RQ is used to investigate the relationship between OSS support tools and the user support they provide. The dependent variable is OSS user support, whereas the independent variables are projected trackers, troubleshooting guidelines, user mailing lists, documentation, and updated versions. We have assumed that OSS user support is depicted by weekly downloads of a project. Let us have a brief discussion of independent variables here.

Project trackers are used to ingmanaging the projects. They may be used to measure and report different things. According to Wikipedia [24], "Troubleshooting is a form of problem-solving, often applied to repair failed products or processes. It is a logical, systematic search for the source of a problem so that it can be solved, and so the product or process can be made operational again." User mailing lists are provided for discussing issues pertaining to installation, configuration, administration, usage, and users own development. Software documentation is a "written text that accompanies computer software. It either explains how it operates or how to use it or may mean different things to people in different roles [25]." Since updated versions are mainly to address security vulnerabilities in a software product, software updates occasionally contain bug fixes and product enhancement.

#### A. Propositions

Five propositions are suggested to examine OSS user support through group project trackers, troubleshooting guidelines, user mailing lists, documentation, and updated versions, which are the five independent variables of the study model. The dependent variable of the model is OSS user support.

In order to carry out the empirical analysis of the research question, we propose the following:

- P1: Project trackers have a positive impact on OSS user support.
- P2: Troubleshooting guidelines are positively related to OSS user support.
- P3: User mailing lists have a positive impact on OSS user support.
- P4: Documentation is positively related to OSS user support.
- P5: Updated versions play a positive role in OSS user support.

The multiple linear regression equation of the model is as follows:

OSS User Support = $c0+c1v1+c2v2+c3v3+c4v4+c5v5$   (1)

Where $c0, c1, c2, c3, c4,$ and $c5$ are the coefficients and $v1, v2, v3, v4,$ and $v5$ are the five independent variables.

### IV. DATA COLLECTION

Data were collected from 100 projects on sourceforge.net, a popular OSS project repository. The dataset covers the ten top-rated projects each from the categories of audio and video, science and engineering, communication, software development, games, graphics, security and utilities, system administration, home and education, and business and enterprise. The maximum weekly downloads of 837,867 were found in the category of software development. Overall, 62 out of 100 studied projects had project trackers. Of the OSS projects that we studied, 73 out of 100 had had troubleshooting guidelines. Only 36 had user mailing lists to discuss issues related to user concerns and requirements about the projects. 54 of these projects happened to have some documentation and 81 out of 100 had versions updated in the year 2016.

### A. Validity Analysis of Measuring Instrument

Validity analysis is conducted in relation to the measuring instruments designed for this research work by utilizing the most common approaches generally used in empirical studies, as presented in Table I below.

TABLE I.    PRINCIPAL COMPONENT ANALYSIS (PCA) OF VARIABLE

| User Support Tools | PCA Eigenvalue |
|---|---|
| Project trackers | 1.31 |
| Troubleshooting guidelines | 1.01 |
| User mailing lists | 1.23 |
| Documentation | 1.05 |
| Updated versions | 1.12 |

The principal component analysis (PCA) [26] was carried out and presented for the five independent variables in Table I. According to Campbel and Fiske [27], convergent validity ensues in a model when the scale items are correlated and have the same direction. Eigenvalues [28] were implied as a reference point, and the construct validity was observed through PCA. In this study, Eigen value-one-criterion was used, also known as Kaiser Criterion [29, 30], which recommends retention of any component having an Eigenvalue greater than unity. The analysis showed that all five variables formed a unit factor, thus establishing the convergent validity as sufficient.

## V.    RESEARCH METHOLOGY

To carry out the analysis of the study model and to investigate the significance of propositions P1, P2, P3, P4, and P5, data analysis activity was carried ouit in three phases. In Phase I, the parametric statistical examination was conducted for the propositions. Pearson correlation coefficients were computed for each of the hypotheses in this test. In Phase II, non-parametric statistical analysis was done by computing Spearman correlation coefficients. Both of the statistical approaches (parametric and non-parametric) were applied to increase the external validity of the study. In Phase III of the study, the propositions were tested using the partial least square (PLS) technique. The technique is especially expedient for small-sample-sized data that suffers from complexity and non-normal distribution [31, 32]. Minitab-17 [33] was used to compute statistical results as presented in Table II below.

The Pearson correlation coefficient and t-test were examined between variables used in propositions P1, P2, P3, P4, and P5. The Pearson correlation coefficient between project trackers and OSS user support (represented by weekly downloads) was found to be positive (0.311) at p = 0.019. The correlation coefficient of -0.014 at p-value = 0.888 was observed between the troubleshooting guidelines and OSS user support in the observed projects, and, hence, P2 was rejected. Proposition P3 was accepted based on the Pearson correlation coefficient (0.321) at p = 0.011 between user mailing lists and the number of weekly downloads. Proposition P4 was rejected based on the Pearson correlation of documentation and user support equal to 0.053 at p-value = 0.601. Finally, the Pearson correlation of updated versions in 2016 and user support was found to be 0.312 at p-value = 0.018. Hence, it was observed and recorded that propositions P1, P3, and P5 were found to be statistically significant. However, propositions P2 and P4 were not supported by the parametric analysis and were, therefore, rejected. Due to the relatively small sample size, non-parametric statistical analysis was also carried out in Phase II. This was done by computing the Spearman correlation coefficients to test the propositions.

TABLE II.    PROPOSITIONS TESTING USING PARAMETRIC AND NON-PARAMETRIC CORRELATION COEFFICIENTS

| Proposition | User Support Tools | Pearson Correlation Coefficient | Spearman correlation coefficient |
|---|---|---|---|
| P1 | Project trackers | 0.311* | 0. 286* |
| P2 | Trouble-shooting guideline | -0.014** | -0.027** |
| P3 | User mailing lists | 0.321* | 0.285* |
| P4 | Documentation | 0.053** | -0.051** |
| P5 | Updated versions | 0.312* | 0.402* |

\* Significant at p < 0.05. ** Insignificant at p > 0.05.

Proposition P1 was found to be statistically significant at p = 0.023 with a Spearman correlation coefficient of 0.286. However, a negative association was observed between troubleshooting guidelines and user support with Spearman rho = -0.027 at p-value = 0.795. The correlation for the user mailing list and user support was found to be significant with Spearman rho = 0.285 p-value = 0.025. For proposition P4, Spearman rho for documentation and user support was observed to be -0.051 at p-value = 0.612. And for the last proposition, Spearman rho for updated versions and user support was observed to be 0.402 at p-value = 0.000. Hence, similar to Phase I, propositions P2 and P4 was rejected as they were not supported by the non-parametric analysis. Propositions P1, P3, and P5, which relate project trackers, user mailing lists, and updated versions to user support were accepted.

In Phase III of proposition testing, the PLS technique was used to do cross-validation of the results, observed in Phase I and Phase II. In PLS, the dependent variable of the study model, i.e., OSS user support, was placed as the response variable with independent variables as predicators.

The test results that contain observed values of path coefficient, $R^2$, and F-ratio, are shown in Table III. Proposition P1 (project trackers – OSS user support) was observed to be significant at p < 0.05 with path coefficient 9909, $R^2$: 0.17 and F-ratio as 22.69. Troubleshooting guidelines have a path coefficient of 28054 with $R^2$: 0.02 and F-ratio of 0.02 and were found to be insignificant at p < 0.05. User mailing lists were observed to have the same direction as proposed in proposition P3 with the path coefficient: 9395, $R^2$: 0.53 and F-ratio: 25.54 at $p < 0.05$. Documentation was found not to be in conformance with the proposition P4 with observed values of path coefficient: 20487, $R^2$: 0.02 and F-ratio: 0.27 at $p > 0.05$. And, finally, updated versions (path coefficient: 1970, $R^2$: 0.41 and F-ratio: 26.45 at $p < 0.05$) were found to be in accordance with P5.

TABLE III. PROPOSITIONS TESTING USING PARTIAL LEAST SQUARE (PLS) REGRESSION

| Proposition | User Support Tools | Path Coefficient | $R^2$ | F- Ratio |
|---|---|---|---|---|
| P1 | Project trackers | 9909 | 0.17 | 22.69* |
| P2 | Troubleshooting guidelines | 28054 | 0.02 | 0.02** |
| P3 | User mailing lists | 9395 | 0.53 | 25.54* |
| P4 | Documentation | 20487 | 0.02 | 0.27** |
| P5 | Updated versions | 1970 | 0.41 | 26.45* |

* Significant at $p < 0.05$. ** Insignificant at $p > 0.05$.

Hence in this phase, as in phase I and phase II, propositions P2 and P4 were not found to be statistically significant at $p < 0.05$.

*A. Research Model Testing*

Equation (1) in Section III depicts the multiple linear regression equation of the research model. The prime purpose of the study was to test empirically whether the selected support tools play a significant role in supporting OSS users. We, thus, carried out regression analysis and computed the values of the model coefficients and their associated directions. OSS user support (the dependent variable) was placed as a response variable and the studied support tools (independent variables) as their predicators. Table IV presents the results of this analysis. The path coefficient of three out of five variables: project trackers, user mailing lists, and updated versions were found to be positive and their t-statistics were also observed to be statistically significant at $P < 0.05$. The path coefficients of troubleshooting guidelines and documentation were found to be negative. Negative t-statistics and $P > 0.05$ made these variables statistically insignificant in the multiple linear regression analysis of the research model. $R^2$ and adjusted $R^2$ of the overall research model are observed as 0.828 and 0.335 with an F-ratio of 11.68 significant at $P < 0.05$.

VI. DISCUSSIONS

Considering user support, OSS is a more perplexing milieu because there are different types of users, both expert, and novice, who belong to different parts of the world and have their own unique experiences and requirements. We have analyzed different contributing support tools in OSS projects and their effect on user support. As stated by Cetin and Gokturk [34], OSS projects can have an increased level of acceptance among end users through measurement and analytical studies. It is highlighted by Hedberg et al. [12] too that understanding user demands, their active contribution, and the context in which software is used are three vital factors that need to receive attention in order to improve OSS designs. In order to have better quality OSS, designers, developers, and user interface experts need to have an active collaboration [35].

There are varied contributors in the OSS environment who share their thoughts through different platforms. Some of these platforms are dedicated and used to actively manage user requirements. In this study, we investigated and analyzed the

TABLE IV. REGRESSION ANALYSIS RESULTS

| User Support Tools | Model Coefficient | Coefficient Value | t-value |
|---|---|---|---|
| Project trackers | $\beta_1$ | 2.81 | 1.45* |
| Troubleshooting guidelines | $\beta_2$ | -1.11 | -0.48** |
| User mailing lists | $\beta_3$ | 4.22 | 2.1* |
| Documentation | $\beta_4$ | -1.01 | -0.47** |
| Updated versions | $\beta_5$ | 1.71 | 0.68* |
| Constant | $\beta_0$ | 1.88 | 0.67* |

* Significant at $p < 0.05$. ** Insignificant at $p > 0.05$.

association between five such platforms, namely group project trackers, troubleshooting guidelines, user mailing lists, documentation, and updated versions and user support.

However, like any other empirical work, this research study is subjected to some limitations. Although we carried out many procedures to enhance the reliability of the data and to reduce the threats to external validity, there are still some limitations. We have used five independent variables to relate to the dependent variable of OSS user support. We realize that there may be other contributing factors that might affect OSS user support besides these five support tools, the scope of this study was limited to examine the role of these specific contributing factors. Another noteworthy limitation of this study is the small sample size in terms of the number of projects. We collected data from 100 projects on sourceforge.net, from the ten top-rated projects in ten different categories. Moreover, OSS user support is depicted by weekly downloads. Since the practitioners in different fields are different, the number of software and the number of downloads could be different.

Understanding the role of contributing factors toward OSS user support through empirical investigation is a challenging project. In this study, we empirically investigated the role of five contributing factors towards user support. The empirical results of the study support the propositions that project trackers, user mailing lists, and updated versions are positively associated with the user support provided by OSS projects. However, no statistical significance could be found for troubleshooting guidelines and documentation, in the phases of parametric, non-parametric, PLS, and multiple regression analysis.

VII. CONCLUSION

In this study, we have examined the user support being provided by different OSS projects. The relationship between user support provided by OSS projects and different user support tools – namely project trackers, troubleshooting guide, user mailing lists, documentation, and updated versions – was studied. The results based on an empirical analysis indicate that project trackers, user mailing lists, and updated versions have a significant role in user support as provided by OSS projects. However, the outcome of our empirical investigation did not support any momentous relationship between user support and documentation, as well as between user support and troubleshooting guidelines as currently being provided by OSS projects. We realize that this work is just a step forward in studying user support by OSS projects. We believe that work should continue to explore more factors and their role toward end-user support in open source software.